\RequirePackage[T1]{fontenc}
\documentclass[journal]{IEEEtran}
\usepackage{amsmath}
\usepackage{amssymb}
\usepackage[numbers,sort&compress]{natbib}
\usepackage{graphicx}
\usepackage{float}
\usepackage{multirow}
\usepackage{xcolor}
\usepackage{placeins}
\usepackage{array}
\newcommand{\rev}[1]{{\color{blue}#1}}
\begin{document}

\title{Frequency Measurement Practices for Inertia Assessment in Inverter-Based Resources--Dominated Power Systems}

\author{Christos M. Nikolakakos,
        Hassan Haes~Alhelou,~\IEEEmembership{Senior Member,~IEEE,}
        and~Nikos~Hatziargyriou,~\IEEEmembership{Life~Fellow,~IEEE}%
\thanks{C. M. Nikolakakos and N. Hatziargyriou are with the School of Electrical and Computer Engineering, National Technical University of Athens, Athens, Greece (e-mail: cnikolakakos@mail.ntua.gr, nh@power.ece.ntua.gr).}%
\thanks{H.H. Alhelou is with the School of Engineering, Massachusetts Institute of Technology (MIT), Cambridge, MA 02139, USA (e-mail: alhelou@ieee.org).}%
}

\maketitle   

\begin{abstract}
\rev{Transmission system operators and protective relays compute the rate of change of frequency (RoCoF) from frequency averaged over tens to hundreds of milliseconds, as standards and grid codes specify. Any response delivered inside that window lowers the measured RoCoF, so the inferred inertia depends on the window and the responses in service. This paper defines the \emph{windowed inertia estimate} and derives an exact identity---the system inertia divided by one minus the fraction of the disturbance energy delivered inside the window---and a closed-form prediction that applies the same estimator to the trajectory of a linear response model. Eleven IEEE 9-bus RMS runs---synchronous, grid-following and grid-forming configurations, disturbance sizes and parameter sweeps---are reproduced by that model, and the identity closes on the measured response powers. Fast frequency response inflates the estimate most: at 500\,ms a grid-following system with 20\% less synchronous inertia reports the same value as the all-synchronous system, the grid-forming system a higher one. On a reduced Greek transmission system the 500\,ms estimate exceeds the inertia by 17--26\%, and a scenario with 10\% less synchronous inertia reports more than the all-synchronous scenario's inertia. Operators can thus compute what their measurement reports.}
\end{abstract}


\section{Introduction}
\IEEEPARstart{P}{ower} systems are undergoing globally a fundamental transformation, driven by the rapid integration of renewable energy sources (RES) such as wind and solar power. Historically, grid stability has relied on large, dispatchable synchronous generators (SGs), which inherently provide the critical stabilizing property of inertia. Stored in the rotating masses of SGs, inertia acts as a natural buffer against sudden imbalances between electricity generation and demand. When disturbances occur---such as a generator trip or a sharp change in load---the kinetic energy stored in these rotating masses is automatically released or absorbed, counteracting the imbalance and limiting the rate of frequency deviation.

The displacement of SGs by inverter-based resources (IBRs) \rev{changes this picture}. IBRs, connected through power electronic converters, do not possess rotating masses directly coupled to the grid, resulting in a measurable decline in physical inertia \cite{10210321}. Consequently, modern power systems experience faster rates of change of frequency (RoCoF) and deeper frequency excursions, which challenge traditional protection systems and increase the risk of instability or cascading outages \cite{hatzi2021}.

Nevertheless, IBRs, through advanced control strategies and proper energy storage, can provide synthetic inertia and fast frequency response (FFR) on timescales far shorter than those of SGs. Unlike synchronous inertia, which is a passive property of mechanical rotation, the inertial response of IBRs is dictated entirely by control design and the available headroom in the converter. As recent experience has shown, the frequency support provided by IBRs, while not identical, can in some cases be comparable or even faster to that of SGs \cite{fang}.

A central challenge lies in how these contributions are assessed in practice. While simulations often capture the instantaneous inertial response and FFR of IBRs, transmission system operators (TSOs) typically measure system frequency through filtered signals, applying techniques originally designed for SG-dominated systems. As a result, the observed response differs significantly from the underlying dynamics: components of FFR may be masked or, conversely, interpreted as inertial behavior. Despite its practical importance, the influence of frequency measurement filters on the perceived contribution of IBRs has received limited attention in the existing literature. This mismatch between simulated dynamics and measured system quantities highlights the critical role of measurement definitions in appreciating the apparent contribution of virtual inertia. Consequently, any accurate assessment of IBR-driven frequency support must explicitly account for the filtering and signal-processing methods employed by TSOs.

While theoretical frameworks for synthetic inertia \cite{DArco2014} and GFM/GFL inverter capabilities \cite{Tayyebi2020,Zhang2021GFM,Poolla2019} are well-established, systematic analysis of how these technologies appear under the filtered RoCoF measurements actually employed by TSOs is not sufficiently explored in current literature. Grid codes specify frequency withstand requirements but leave measurement methodologies undefined \cite{ncrfg_2016,entsoe_rocof_igd_2018}, creating potential inconsistencies in how IBR frequency support services are assessed and valued.

\rev{This paper addresses this gap analytically and through simulation. Its contributions are: (i) a framework for the \emph{windowed inertia estimate} $\hat{H}(T_w)$---the inertia a RoCoF measurement with averaging window $T_w$ reports---consisting of an exact identity in terms of the delivered fractions of the frequency responses and a unified prediction that applies the same estimator to a modelled frequency trajectory---in closed form for a linear response model in which governor lags, the grid-following measurement delay, power-control lag and filtered derivative response all enter as first-order lags---each response entering through the part of it delivered inside the window; (ii) its validation on the IEEE 9-bus benchmark: eleven RMS simulation runs spanning SG-only, GFL and GFM configurations, disturbance sizes, governor and converter parameter sweeps, each compared with the model evaluated from declared settings and the identified governor and power-loop equivalents, and the application of the estimator to three scenarios of a reduced model of the Greek transmission system, which quantifies the window dependence of the estimate across scenarios with different synchronous inertias; and (iii) the placement of standard reference RoCoF filters (IEC/IEEE 60255-118-1:2018) and of a model-based correction within the same framework, with their limits stated. The unfiltered estimate returned the reference inertia within 2.5\% in every run, while windows of 100--500\,ms raised it by factors of up to 1.36 (SG), 1.66 (GFL) and 1.71 (GFM), re-classifying fast frequency response as inertia.} These findings have direct implications for grid code design, market mechanisms and system planning in IBR-dominated networks.

\section{Frequency Stability in Power Systems}

\subsection{Traditional Frequency Response Framework}

Frequency response following a power imbalance is traditionally divided into three stages: (i) the inertial response of SGs, which instantaneously counteracts power imbalances through stored kinetic energy; (ii) the primary frequency response, provided by governor action and load damping; and (iii) secondary control, which restores frequency to its nominal value via automatic generation control. In systems dominated by SGs, these mechanisms are well-defined and inherently linked to the physical characteristics of rotating machines.

IBRs, by contrast, do not inherently provide inertial response. Their contribution to frequency stability arises through control actions that can provide synthetic inertia. These services are not constrained by mechanical inertia but by converter and control design, which allows for faster responses and potentially larger effective droop settings compared with SGs \cite{Lasseter2020}.  The equivalence between virtual synchronous machines and frequency-droop control strategies, providing a unified framework for analyzing IBR frequency support is well documented in literature. \cite{DArco2014}.

\subsection{Inertia-Frequency Relationship in Power Systems}

The system dynamics can be described by the classical swing equation:
\begin{equation}
2H\,\frac{df(t)}{dt} = P_m(t) - P_e(t)
\label{eq:swing}
\end{equation}
where $H$ denotes the inertia constant (in seconds), $f$ the system frequency, $P_m$ the mechanical input power, and $P_e$ the electrical output power (all expressed on a consistent per-unit base). The instantaneous RoCoF is then given by:
\begin{equation}
\text{RoCoF}(t) = \frac{df(t)}{dt} = \frac{1}{2H}\bigl(P_m(t)-P_e(t)\bigr) = \frac{\Delta P(t)}{2H}
\label{eq:rocof}
\end{equation}

The total system inertia is considered to originate from SGs and grid-forming (GFM) inverters. Grid-following (GFL) inverters are excluded from the \rev{reference inertia} $H$, \rev{since their response is delivered behind PLL synchronization, measurement delays and fault-ride-through logic, preventing an instantaneous contribution at the disturbance instant} \cite{Poolla2019}.

This simplified relationship provides a clear analytical link between system inertia and frequency dynamics; however, it does not account for the effect of measurement processes. In practice, the RoCoF observed by TSOs is shaped by filtering and signal-processing requirements imposed by grid codes, which modify the apparent response and, consequently, the inferred inertia.

\subsection{\rev{Low-Inertia Phenomena, Time Scales and Distributed Resources}}

\rev{The transition towards low-inertia systems raises the initial RoCoF, deepens frequency nadirs, lowers the quasi-steady-state frequency and increases the sensitivity of frequency trajectories to protection actions and discrete events \cite{Milano2018RoCoF}. These phenomena have motivated system-wide metrics---minimum inertia requirements, \emph{effective} system inertia including synthetic-inertia and FFR contributions, and frequency-security indicators based on nadir, RoCoF and recovery time \cite{ENTSOE_InertiaRoCoF_2020,AEMO_InertiaRequirements_2024}---all of which presuppose a consistent measurement of frequency and RoCoF and therefore inherit a dependence on the measurement window and filtering adopted by the TSO. In terms of time scales, synchronous inertia acts within 0--0.5\,s, control-based synthetic inertia and FFR within 50--500\,ms, primary frequency response within 1--10\,s, and secondary control beyond tens of seconds: GFM inverters can be designed to participate in the inertial window, whereas GFL inverters act as fast FFR providers behind PLL-induced delays \cite{Poolla2019,Zhang2021GFM}---distinctions that filtered frequency measurements blur. Distributed resources add a further layer: RoCoF- and frequency-sensitive protection of distributed generation shaped the South Australia 2016 and Great Britain 2019 events, illustrating that measurement windows and protection settings at the distribution level can affect system-wide frequency stability \cite{SA_Event_2016,UK_Event_2019}.}

\subsection{\rev{Measurement Standards, Grid Codes and RoCoF Windows}}
\label{sec:freq-meas-standards}

Accurate estimation of frequency and RoCoF is central to the interoperability of grid-forming and grid-following inverters. However, grid codes rarely prescribe the exact measurement algorithm; instead they define operating envelopes (frequency ranges, RoCoF withstand limits, ride-through zones) and leave measurement dynamics to international standards and manufacturers. This section summarises the measurement standards and the grid-code practices that set the windows studied here, before formalising the effect of the measurement window on \rev{the reported inertia through the windowed inertia estimate $\hat{H}(T_w)$}.

\subsubsection{International Measurement Standards}

The global reference for dynamic frequency and RoCoF measurement is the IEC/IEEE~60255-118-1:2018 synchrophasor standard \cite{iec60255_2018}. It defines the measurement of synchrophasors, frequency and RoCoF under both steady-state and dynamic conditions, introduces performance classes (P-class and M-class), and specifies test waveforms such as frequency ramps, steps and oscillations. This standard harmonises and supersedes the measurement component of the earlier IEEE C37.118.1 specification, which remains widely referenced in PMU deployments.

In contrast, power-quality instruments typically comply with IEC~61000-4-30 Class~A, which measures the power frequency as a cycle count over a 10\,s interval and aggregates the RMS-based quantities over 10/12-cycle and 150/180-cycle intervals \cite{iec61000_2015}. While suitable for monitoring and compliance reporting, these windows introduce significant latency and are therefore unsuitable for control loops requiring sub-second response (e.g., inertia emulation or FFR), as highlighted in recent RoCoF estimation surveys \cite{freq_rocof_survey_2023}.

Recent metrological research has quantified the accuracy requirements for RoCoF measurements in different applications \cite{Rietveld2020RoCoF}. For under-frequency load shedding (UFLS), an accuracy of 0.01\,Hz/s with latency under 50\,ms is considered ideal, though practical implementations typically achieve 0.1\,Hz/s accuracy. Fast frequency response and synthetic inertia applications require RoCoF measurements with latency under 100\,ms and accuracy better than 0.1\,Hz/s. Loss-of-mains (LOM) protection can tolerate latencies up to 250\,ms but requires consistent accuracy to prevent nuisance tripping. The relaxed IEEE standard specifies 0.4\,Hz/s accuracy under modulated signal conditions, acknowledging the practical challenges of real-time RoCoF estimation in disturbed grid conditions.

\subsubsection{\rev{Grid-Code Practice and the Windows Studied}}
\label{sec:rocof_windows}

The European Network Codes---the Requirements for Generators (NC~RfG), NC~DCC and NC~HVDC---define frequency operating ranges, mandatory frequency response modes (LFSM-O/U) and RoCoF withstand obligations for generating modules \cite{ncrfg_2016}, but do not specify the measurement method for frequency or RoCoF; the ENTSO-E Implementation Guidance Document treats RoCoF as a system-level quantity and leaves the estimator, the window length and the filtering to TSOs and manufacturers \cite{entsoe_rocof_igd_2018}. Where a reference window is stated it is long: ENTSO-E guidance and the Irish and British studies converge on a RoCoF measured over a 500\,ms rolling window for adequacy assessment and loss-of-mains (LOM) protection \cite{entsoe_rocof_igd_2018,eirgrid_rocof_2016,ngeso_lom_2019,ENTSOE_InertiaRoCoF_2020}, and AEMO's inertia framework and frequency-control reviews use RoCoF calculated over a nominal 500\,ms window when determining minimum inertia levels and FFR requirements \cite{aemo_international_review_2020,aemo_inertia_review_2021,AEMO_InertiaRequirements_2024}, with the Frequency Operating Standard defining the reference frequency measurement used for compliance \cite{aemc_fos_2020} and distribution protection practice using 200--500\,ms \cite{aemo_protection_guide_2020}. In North America, NERC PRC-024-4 requires the frequency used by protective relays to be calculated over a time window---trip settings based on an instantaneously calculated frequency are not permissible, although instantaneous trip thresholds outside the specified boundaries are---and gives 3--6 cycles (50--100\,ms at 60\,Hz) as the typical window length (Attachment~1, footnote~10) \cite{nerc_prc024_2024}, while PMU-based monitoring and system studies work with 40--100\,ms windows \cite{iec60255_2018,freq_rocof_survey_2023,aemo_international_review_2020}. In every case the withstand requirement is specified and the measurement is left to the estimator, so a protection relay and an operational study see different RoCoF values for the same event.

\rev{The measurement chains studied in this paper---the unfiltered slope and 100, 300 and 500\,ms moving averages---span this range: the short end corresponds to PMU-based monitoring and to the NERC averaging requirement, the long end to the 500\,ms reference used in Europe and Australia for adequacy assessment and inertia requirements. Section~\ref{sec:pmu_reference} additionally applies the reference filters of IEC/IEEE 60255-118-1 to the same records. The same standard's ROCOF response-time limits (120\,ms for P class; 280--560\,ms for M class at 50/25 frames/s, Table~9), its reporting latency of up to $7/F_s$ (Table~10) and the multi-report averaging applied by ROCOF-based applications place decisions built on such measurements on the same 100--500\,ms band; the standard notes that latency includes ``the window over which data is gathered''. Response time, latency and averaging window are distinct quantities and are not equated here.}

\section{\rev{Inverter-Based Resource Response Models}}

\subsection{Grid-Forming Inverter Characteristics}

GFM inverters implementing Virtual Synchronous Generator (VSG) control provide synthetic inertia by emulating the swing equation dynamics of a synchronous machine. The VSG control law can be expressed as:
\begin{equation}
P_{\text{VSG}} = P_{\text{ref}} - 2H_v S_{\text{rated}}\frac{df}{dt} - D_p(f - f_0)
\label{eq:vsg_control}
\end{equation}
where $H_v$ represents the virtual inertia constant, $S_{\text{rated}}$ is the rated power, and $D_p$ is the damping coefficient. The ability of GFM inverters to behave as voltage sources behind an impedance enables immediate power injection in response to frequency deviations, \rev{which distinguishes} them from GFL inverters \cite{Tayyebi2020}.

Experimental validation confirms that VSG implementations accurately replicate synchronous machine dynamics \cite{Abuagreb2020VSG}. When viewed through a moving-average filter, \rev{the windowed inertia estimate} can exceed the designed virtual inertia parameter because any power injection within the measurement window contributes to the perceived response \cite{Poolla2019}\rev{---in the terms of Section~\ref{sec:framework} below, $D_p$ acts on the actual frequency and belongs to the undelayed aggregate $D$, and the primary response of the remaining machines enters through its delivered fraction at long windows}.

\subsection{Grid-Following Inverter Behavior}

GFL inverters, \rev{whose response is delivered behind PLL-based synchronization and therefore does not act at the disturbance instant}, can exhibit \rev{an inflated windowed inertia estimate} when their FFR is filtered. The PLL introduces a measurement delay that can be approximated by a first-order lag \cite{Poolla2019}:
\begin{equation}
f_{\text{PLL}}(s) = \frac{f_{\text{grid}}(s)}{1 + s T_{\text{PLL}}}
\label{eq:pll_response}
\end{equation}
where $T_{\text{PLL}}$ is the PLL time constant, typically in the range of 20--100\,ms depending on bandwidth settings. \rev{In the implementation studied here the SRF-PLL has a first-order-equivalent lag that is small against the other terms; the dominant measurement and delivery effects are an explicit 40\,ms frequency-measurement delay, a $\pm 50$\,mHz activation deadband, and an equivalent delivery lag of $\approx 0.14$\,s standing for the power-control chain (Section~\ref{sec:provenance}).}

GFL inverters with FFR capability inject power, with certain delay, proportional to the measured frequency deviation:
\begin{equation}
P_{\text{FFR}} = K_f (f_0 - f_{\text{PLL}})
\label{eq:ffr_power}
\end{equation}
where $K_f$ is the frequency droop gain\rev{, expressed in per-unit power (converter base) per per-unit frequency}. When this response is observed through a measurement filter of window $T_w$, the FFR contribution becomes indistinguishable from inertial response in the filtered signal. \rev{The windowed contribution} increases with longer measurement windows and higher droop gains \cite{Poolla2019,fang}.

\rev{Swing-equation analysis shows} that while GFM/SG inertia directly limits initial RoCoF, the equivalent GFL inertia term does \emph{not} constrain initial frequency change \cite{Ducoin2023GFL}. \rev{For the parameters studied here ($T_d = 40$\,ms, $K_f = 20$\,pu on the converter base, $T_w$ up to 500\,ms), Section~\ref{sec:results_ieee9} measures this directly: the converter contributes nothing to the initial RoCoF---its implemented RoCoF-emulation branch included---while the 500\,ms window re-classifies its FFR into an overstatement of the system's physical inertia. It is also mathematically proven in the following section, that any frequency response behind a delay cannot limit initial RoCoF}

\section{\rev{The Windowed Inertia Estimate}}

\subsection{\rev{Filtered RoCoF}}

In practice, RoCoF is not measured as an instantaneous derivative but is obtained from frequency estimates filtered over a finite time window. A simple and widely used representation is the moving-average (MA) derivative \cite{freq_rocof_survey_2023}:
\begin{equation}
\text{RoCoF}_{\text{measured}}(t) = \frac{f(t) - f(t-T_w)}{T_w}
\label{eq:rocof_measured}
\end{equation}
where $T_w$ represents the measurement window length. This expression captures the essence of RoCoF measurements used in protection relays and adequacy studies, where $T_w$ typically ranges from tens of milliseconds (PMU-based monitoring) to several hundred milliseconds (LOM protection and compliance monitoring) \cite{entsoe_rocof_igd_2018,eirgrid_rocof_2016,freq_rocof_survey_2023}.

For a step power disturbance $\Delta P(t) = \Delta P \cdot u(t)$, where $u(t)$ is the unit step function, the magnitude of the filtered RoCoF can be derived analytically in the inertia-only case, in which the RoCoF is the constant $-\Delta P/2H$ after the event. Averaging its magnitude over the moving window yields:
\begin{equation}
|\text{RoCoF}_{\text{filt}}(t)| =
\begin{cases}
0, & t < 0,\\[4pt]
\displaystyle \frac{\Delta P}{2H T_w}\, t, & 0 \le t < T_w,\\[6pt]
\displaystyle \frac{\Delta P}{2H}, & t \ge T_w.
\end{cases}
\label{eq:rocof_filt_piecewise}
\end{equation}
This inertia-only expression reveals two observations: (i) the filtered RoCoF increases linearly during the measurement window rather than exhibiting the step response of the true RoCoF, and (ii) the filtered value converges to the theoretical instantaneous value only after the full window duration $T_w$ has elapsed. Consequently, any frequency support delivered during the interval $0 \le t < T_w$ contributes to reducing the \emph{measured} RoCoF, regardless of whether that support originates from physical inertia or fast-acting control.

The filtering \rev{changes} the observed dynamics. For a step change in power, the actual RoCoF exhibits an instantaneous jump followed by exponential decay, while the measured RoCoF shows a gradual rise and fall shaped by the filter characteristics. \rev{Fast} frequency response delivered within the measurement window becomes indistinguishable from physical inertia in the filtered signal.

Recent ENTSO-E and AEMO documents explicitly acknowledge this by introducing \emph{effective} or \emph{measured} inertia that differs from physical inertia \cite{ENTSOE_InertiaRoCoF_2020,ENTSOE_GFM_2024,AEMO_InertiaRequirements_2024}. Since system inertia must be estimated from post-disturbance frequency response using measurement windows of approximately 500\,ms, physical inertia and FFR contributions become operationally indistinguishable. \rev{Section~\ref{sec:framework} makes this dependence quantitative: the windowed inertia estimate is written as the inertia divided by one minus the delivered response fractions, Eq.~\eqref{eq:hhat_identity}, and computed from the response constants of the units in service by Eqs.~\eqref{eq:hhat_general}--\eqref{eq:hhat_expansion}.}

\subsection{\rev{Definition and Analytical Framework}}
\label{sec:framework}

\rev{The inertia $H$ is the energy-based aggregate on the common system base, $H=\sum_i H_i S_{gn,i}/S_B$, comprising the kinetic energy of the synchronous machines and the programmed inertia constant of GFM virtual rotors (a control coefficient supplied by the converter's energy source, not rotating mass; it is counted in $H$ because it acts on the derivative of the actual frequency). The converter then contributes only its damping $D_p$ to the responses. The frequency signal $\Delta f$ is the centre of inertia (COI) of the synchronous machines in service (in the GFM case the converter's internal frequency is taken equal to it, Assumption~(ii)), $\Delta f = \sum_i E_i \Delta\omega_i / \sum_i E_i$ with $E_i = H_i S_{gn,i}$; the window $T_w$ is a property of the estimator only, applied in post-processing; and the effective disturbance $\Delta P_{\mathrm{eff}}$ is measured per run.}
\rev{Assumptions: (i) responses from zero initial conditions, linear and unsaturated wherever the closed-form evaluation and the planning form are used (the estimator and the identity do not need linearity; the GFL droop deadband is outside the linear model and its effect is quantified in Section~\ref{sec:evaluation}); (ii) the GFM behaves as a voltage source whose programmed inertia is lumped into $H$ and whose damping acts on the actual frequency; (iii) the PLL of a GFL converter is represented by its measurement delay, modelled as a first-order lag of the same time constant, and its power-loop lag; (iv) the load-side response after the event instant is neglected in the reduced model, its measured size being stated in Section~\ref{sec:results_ieee9} (a frequency-proportional load damping could be added to $D$ where a system requires it); (v) $\Delta f$ is the centre-of-inertia frequency; (vi) windowed statements concern windows shorter than the time to the frequency nadir.}

\rev{The estimator \eqref{eq:hhat_def} defines the windowed inertia estimate from a RoCoF record, measured or modelled. The energy identity \eqref{eq:hhat_identity} writes it in terms of the delivered response energies; it holds for any response model and is used here with measured response powers to attribute a measured bias. The response model \eqref{eq:model_td}, one closed loop with each response as its own state, is what the predictions are computed from: solving it and applying the estimator to its trajectory is the model prediction \eqref{eq:hhat_general}, whose closed form shows how the estimate is fixed by the closed-loop modes. The planning form \eqref{eq:hhat_expansion} then shows what each response does to the estimate---an undelayed response adds a term linear in the window, a first-order-lag one a term quadratic in the window set by its build-up rate---and gives a calculation for planning.}

\rev{The responses enter the swing equation \eqref{eq:swing} through their own dynamics. In the Laplace domain, with zero initial conditions and the GFL droop deadband neglected (its effect is quantified in Section~\ref{sec:evaluation}), a step deficit $\Delta P$, unlagged damping $D$ and $D_p$, governor branches $K_i$ behind lags $T_i$ (sharing the droop $K_g$), and a GFL droop $K_f$ and RoCoF-emulation gain $2H_{GFL}$ that both act on a frequency measured behind a delay $\theta$, represented by a first-order lag of that time constant, and are delivered through the converter's power-control lag $T_f$, the emulation also through its RoCoF filter $T_r$, the closed loop reads}
\begin{equation}
\rev{\begin{aligned}
\Delta f(s) &= -\frac{\Delta P}{s\,Q(s)},\\
Q(s) &= 2H\,s + D + D_p + \sum_i\frac{K_i}{1+sT_i}\\
&\quad + \frac{K_f}{(1+s\theta)(1+sT_f)}\\
&\quad + \frac{2H_{GFL}\,s}{(1+s\theta)(1+sT_r)(1+sT_f)},
\end{aligned}}
\label{eq:heff}
\end{equation}
\rev{where $H = H_{SG}+H_{GFM}$ comprises the synchronous machines and the GFM virtual rotor (a voltage source, unlagged). The initial RoCoF follows from the initial-value theorem: as $s \to \infty$ the deviation-driven terms are bounded and the lagged terms vanish, so every other term of $Q(s)$ is negligible against $2Hs$ and}
\begin{equation}
\rev{\mathrm{RoCoF}(0^{+}) = \lim_{s\to\infty} s\left[s\,\Delta f(s)\right] = -\frac{\Delta P}{2\,(H_{SG}+H_{GFM})},}
\label{eq:rocof0}
\end{equation}
\rev{independently of $K_f$, $K_g$, $D$, $D_p$, $H_{GFL}$, $\theta$, $T_f$ and $T_r$: only unlagged derivative action survives at an instant when the frequency deviation is still zero and every measurement still reads nothing. Were the emulation delivered without delay or lag ($\theta = 0$, $T_r = T_f = 0$), its term would survive and the limit would read $-\Delta P/2(H+H_{GFL})$ --- the dividing line between the technologies is the measurement and delivery chain, not the control law. This is the mathematical content of the GFL exclusion from $H$ asserted in Section~II.}

\rev{The \emph{windowed inertia estimate} is the inertia the filtered measurement \eqref{eq:rocof_measured} reports,}
\begin{equation}
\rev{\hat{H}(T_w) \;=\; \frac{\Delta P_{\mathrm{eff}}}{2\,\displaystyle\max_t \left| \mathrm{RoCoF}_{\mathrm{measured}}(t) \right|},}
\label{eq:hhat_def}
\end{equation}
\rev{If the initial frequency decline has non-increasing slope magnitude and the window ends before the nadir, the window that starts at the event maximises the averaged slope over that initial decline: a window straddling the event averages pre-event zeros into the fall, and a window starting later averages a smaller slope. Equality with the maximum over the entire record additionally requires that no later window---during the recovery, for instance---has a larger averaged slope; neither property is guaranteed by the loop \eqref{eq:heff}, whose poles may be complex. Both were therefore checked on every record by searching the full recorded interval for the maximising window, which ended at $t_0+T_w$ in every run reported in Section~\ref{sec:results_ieee9}. $\hat{H}(T_w)$ is a measurement construct, not a system property.}

\rev{The responses on the right-hand side of the swing equation \eqref{eq:swing} are classified by their driving signal and their delivery dynamics, not by the service name or the converter architecture. \par  An \emph{undelayed proportional} response is represented by $p = K\,|\Delta f|$, with no lag or dead time between the frequency and the delivered power in the adopted model. The modelled load damping, the machine damping, and any controller branch of this form---here the proportional damping $D_p$ of the grid-forming converter, which as a voltage source responds to the actual frequency (Assumption~(ii)). Their gains sum to one aggregate, written $D$ for continuity with \eqref{eq:heff}: $D = D_{\mathrm{load}} + D_{\mathrm{machine}} + \sum_{j} K_j$ over such branches. ``Undelayed'' does not mean a power step at the event: with $\Delta f(0) = 0$ the power starts from zero and grows with the deviation. \par  A \emph{lagged} response delivers its power through a time constant $T_i$, $T_i\,\dot{p}_i + p_i = K_i\,|\Delta f|$, either because the plant does so on the actual frequency---the governors, through turbine and reheat dynamics---or because the branch acts on a \emph{measured} frequency behind a PLL, a frame delay and a power loop, as the grid-following droop and virtual inertia do; the latter are, in addition, invisible at $t = 0^+$ by \eqref{eq:rocof0}. The measurement delay $\theta$ of the studied GFL controller is represented by a first-order lag of the same time constant, so that every response in the model is a gain or a chain of first-order lags and the whole loop is solved in closed form; the droop deadband of the simulated controller is not represented, and Section~\ref{sec:evaluation} gives the size of both simplifications. \par ``Lagged'' below means any such branch; governor lags are included in the closed form. Ideal derivative action is accounted for in $H$ under the stated convention and never also among the responses. } \par
\rev{For the systems studied, the classes are populated as follows: the undelayed aggregate collects the machine damping and, in the GFM case, $D_p$ (no load-damping term is carried, Section~\ref{sec:results_ieee9}); the governors form two lagged branches on the actual frequency, a turbine branch with time constant $T_g$ and a reheat branch with $T_R$, sharing the declared droop; the GFL droop is a lagged branch on the measured frequency, behind the measurement-delay lag; and the GFL RoCoF-emulation branch is derivative action on the measured frequency, the branch $v$ of \eqref{eq:model_td}, which has no term in the planning form, while the GFM virtual rotor is counted in $H$ and its derivative component is excluded from the responses. The parameter values and their sources are given in Section~\ref{sec:provenance} and Table~\ref{tab:model_inputs}. \par  Let $p_X(t)$ be the power delivered by response class $X \in \{u,\,g,\,m\}$---undelayed proportional; lagged on the actual frequency (governors); lagged on a measured frequency (grid-following). Integrating the swing equation \eqref{eq:swing} over the window that starts at the event, with $\Delta f(0)=0$ and a constant deficit $\Delta P_{\mathrm{eff}}$, gives an energy balance:}
\begin{equation}
\rev{2H\,|\Delta f(T_w)| \;=\; \Delta P_{\mathrm{eff}}\,T_w \;-\; \sum_X \int_{0}^{T_w} p_X(t)\,\mathrm{d}t ,}
\label{eq:energy_balance}
\end{equation}
\rev{in which $\Delta P_{\mathrm{eff}}\,T_w$ is the energy the disturbance has removed during the window and each integral is the energy a response has supplied. Dividing by $\Delta P_{\mathrm{eff}}\,T_w$ expresses every term as a share of the disturbance energy; the share supplied by response group $X$ is its \emph{delivered fraction},}
\begin{equation}
\rev{\rho_X(T_w) \;=\; \frac{1}{\Delta P_{\mathrm{eff}}\, T_w} \int_{0}^{T_w} p_X(t)\, \mathrm{d}t ,}
\label{eq:rho_def}
\end{equation}
\rev{equal to one for a response that had covered the whole deficit throughout the window and to zero for one that has not yet acted. On the left, $|\Delta f(T_w)|/T_w$ is the moving-average RoCoF of the window that starts at the event, which equals the maximum in \eqref{eq:hhat_def} when the stated alignment condition holds, so $2H\,|\Delta f(T_w)|/(\Delta P_{\mathrm{eff}}\,T_w) = H/\hat{H}(T_w)$. The balance therefore reads $H/\hat{H} = 1 - \sum_X \rho_X$, that is, the exact identity}
\begin{equation}
\rev{\hat{H}(T_w) \;=\; \frac{H}{1-\rho_u(T_w)-\rho_g(T_w)-\rho_m(T_w)},}
\label{eq:hhat_identity}
\end{equation}
\rev{valid for any disturbance size, any set of units in service and any delivery dynamics, linear or not, within the aggregate swing balance \eqref{eq:swing}. The disturbance magnitude cancels from the identity itself; $\hat{H}(T_w)$ is independent of $\Delta P$ only when the delivered fractions are, which holds for linear responses from zero initial conditions and fails where a deadband or limiter acts. The delivered response energy sets the inflation through the identity: the estimate exceeds $H$ by $H\sum\rho/(1-\sum\rho)$. As $T_w \to 0$ each fraction tends to $p_X(0^+)/\Delta P_{\mathrm{eff}}$; under the stated convention, with ideal derivative action counted in $H$ and every remaining response starting from zero, all fractions vanish and the unfiltered estimate returns $H$, recovering \eqref{eq:rocof0}---in the GFM case this reference includes the programmed virtual-inertia coefficient. This unfiltered value is the reference against which the windowed estimates are read. Because \eqref{eq:hhat_identity} requires the delivered powers, it is the form used after an event to attribute the bias to the responses.}

\rev{The response model is one set of time-domain equations for every configuration studied. With $x(t) = -\Delta f(t)$ the frequency drop, positive during the initial fall, $x_m$ the drop as the GFL controller measures it, $p_{g,i}$ the governor branches, $p_f$ the GFL droop power and $v$ the GFL RoCoF-emulation power, all on $S_B$,}
\begin{equation}
\rev{\begin{aligned}
2H\,\dot{x} &= \Delta P - D\,x - \textstyle\sum_i p_{g,i} - p_f - v,\\
T_i\,\dot{p}_{g,i} + p_{g,i} &= K_i\,x,\\
\theta\,\dot{x}_m + x_m &= x,\\
T_f\,\dot{p}_f + p_f &= K_f\,x_m,\\
T_r\,\dot{r} + r &= 2H_{GFL}\,\dot{x}_m,\\
T_f\,\dot{v} + v &= r,
\end{aligned}}
\label{eq:model_td}
\end{equation}
\rev{from zero initial conditions, with $D$ the undelayed aggregate (which includes $D_p$ where a GFM is present) and $\theta$ the GFL measurement delay, represented by the first-order lag of the third line; the droop deadband is not represented. The emulation command passes through the same power-control lag $T_f$ as the droop because, in the implemented controller, the two commands are summed into one frequency-response signal ahead of the converter's power loop (Section~\ref{sec:evaluation}); a branch that a configuration does not contain has zero gain.  \par  Transforming \eqref{eq:model_td} gives $X(s) = \Delta P/(s\,Q(s))$ with $Q(s)$ as in \eqref{eq:heff}, $D_p$ absorbed in $D$. Every response is an undelayed gain or a chain of first-order lags, so the closed loop has a finite number of poles and $x(t)$ is a sum of exponentials, the closed form derived below, for all three configurations alike (explicit time-stepping of \eqref{eq:model_td} reproduces every closed-form value of Table~\ref{tab:closedform_bracket} to within $0.01$~s).}

\rev{Before an event the same quantity follows from the model parameters alone, in two steps: the solution of \eqref{eq:model_td} gives the frequency drop at the end of the window, and that drop is substituted into the estimator \eqref{eq:hhat_def}; this is all that a prediction needs, and what follows makes the result readable rather than only computable. Write $g(t) = x(t)/\Delta P$ for the drop per unit of disturbance; for the linear model $g$ does not depend on $\Delta P$, and on the window starting at the event the estimator \eqref{eq:hhat_def} reduces to $T_w/2g(T_w)$, the first line of \eqref{eq:hhat_general} below. The rest of the block evaluates $g(T_w)$ for the model.}

\rev{For the model, multiplying numerator and denominator by the product $M(s)$ of all lag factors---$(1+sT_i)$ for the governor branches and, where a GFL is present, $(1+s\theta)$, $(1+sT_f)$ and $(1+sT_r)$---clears the fractions and turns the denominator into the polynomial $P(s) = s\,Q(s)\,M(s)$. Its root at the origin carries the constant part of the step response; its remaining roots $s_k$ are the closed-loop poles, each governing one transient mode---a real pole a decaying exponential, a complex pair a damped oscillation. Partial fractions over these roots give $g(t) = Q(0)^{-1} + \sum_k r_k e^{s_k t}$, so that the model prediction is}
\begin{equation}
\rev{\begin{aligned}
\hat{H}(T_w) &= \frac{T_w}{2\,|g(T_w)|},\\
g(T_w) &= Q(0)^{-1} + \sum_k r_k\, e^{s_k T_w},\\
r_k &= \frac{M(s_k)}{P'(s_k)},
\end{aligned}}
\label{eq:hhat_general}
\end{equation}
\rev{The first line is the common prediction rule---the estimator \eqref{eq:hhat_def} applied to a predicted trajectory, under the alignment condition stated after \eqref{eq:hhat_def}; the remaining lines give its closed-form evaluation for every configuration studied and expose the structure: the estimate is set by the closed-loop poles $s_k$ and their coefficients $r_k$. Here, $Q(0) = D + \sum_i K_i + K_f$ is the total static gain (the emulation branch has none) and $r_k$ the coefficient of mode $k$; $\Delta P$ has cancelled because $g$ is independent of it. Two identities check a computation and tie it to the earlier results: $\sum_k r_k = -1/Q(0)$, since $x(0) = 0$, and $\sum_k r_k s_k = 1/2H$, the initial RoCoF of \eqref{eq:rocof0}. }
\rev{The prediction \eqref{eq:hhat_general} gives a number; it does not by itself say which response is responsible for how much of it, or how that share scales with the window and the parameters. A planning form that answers this, and needs no root finding, follows from the identity \eqref{eq:hhat_identity} by evaluating each delivered fraction \eqref{eq:rho_def} on the initial trajectory of \eqref{eq:rocof0}, before the responses have bent it. At the event every response power is still zero, so \eqref{eq:swing} reduces to $2H\,\mathrm{d}|\Delta f|/\mathrm{d}t = \Delta P_{\mathrm{eff}}$---the initial RoCoF of \eqref{eq:rocof0}---and with $\Delta f(0)=0$,}
\begin{equation}
\rev{|\Delta f(t)| \approx \frac{\Delta P_{\mathrm{eff}}}{2H}\,t, \qquad 0 \le t \le T_w .}
\label{eq:ramp}
\end{equation}
\rev{\emph{Undelayed proportional response:} The aggregate delivers $p_u = D\,|\Delta f|$. Inserting \eqref{eq:ramp} in \eqref{eq:rho_def},}
\begin{equation}
\rev{\begin{aligned}
\rho_u &= \frac{1}{\Delta P_{\mathrm{eff}}\,T_w}\int_0^{T_w} D\,|\Delta f(t)|\,\mathrm{d}t \\
       &\approx \frac{D}{\Delta P_{\mathrm{eff}}\,T_w}\cdot\frac{\Delta P_{\mathrm{eff}}}{2H}\int_0^{T_w} t\,\mathrm{d}t \\
       &= \frac{D}{2H\,T_w}\cdot\frac{T_w^{2}}{2} = \frac{D\,T_w}{4H},
\end{aligned}}
\label{eq:rho_ramp}
\end{equation}
\rev{\emph{Lagged response:} A response of gain $K_i$ behind the lag $T_i$ obeys $T_i\,\dot{p}_i + p_i = K_i\,|\Delta f|$ with $p_i(0)=0$. At the event $|\Delta f| = 0$, so $\dot{p}_i(0) = 0$ as well; differentiating the lag equation once, $T_i\,\ddot{p}_i + \dot{p}_i = K_i\,\mathrm{d}|\Delta f|/\mathrm{d}t$, gives $\ddot{p}_i(0) = K_i\,\Delta P_{\mathrm{eff}}/(2H\,T_i)$. The response therefore starts as the parabola}
\begin{equation}
\rev{p_i(t) \approx \tfrac{1}{2}\ddot{p}_i(0)\,t^2 = \frac{K_i\,\Delta P_{\mathrm{eff}}}{2H\,T_i}\cdot\frac{t^{2}}{2},}
\label{eq:p_ramp}
\end{equation}
\rev{whose curvature is set by $K_i/T_i$, the rate at which the response's power starts to rise per unit of frequency deviation ($T_i$ is the delivery time constant of Section~\ref{sec:provenance}: the turbine and reheat constants for the governor branches; the GFL droop and filtered emulation powers, behind two or more lags, both start at order $t^3$ and are outside this form). Inserting \eqref{eq:p_ramp} in \eqref{eq:rho_def},}
\begin{equation}
\rev{\begin{aligned}
\rho_i &= \frac{1}{\Delta P_{\mathrm{eff}}\,T_w}\int_0^{T_w} p_i(t)\,\mathrm{d}t \\
       &\approx \frac{K_i}{2H\,T_i\,T_w}\int_0^{T_w}\frac{t^{2}}{2}\,\mathrm{d}t \\
       &= \frac{K_i}{2H\,T_i\,T_w}\cdot\frac{T_w^{3}}{6} = \frac{K_i\,T_w^{2}}{12\,H\,T_i}.
\end{aligned}}
\label{eq:rhoi_ramp}
\end{equation}
\rev{Equations~\eqref{eq:p_ramp} and~\eqref{eq:rhoi_ramp} retain the leading short-time terms, which requires $t$ small against $T_i$. For small delivered fractions, $H/(1-\sum\rho) \approx H(1+\sum\rho)$, giving}
\begin{equation}
\rev{\begin{aligned}
\hat{H}(T_w) &\approx H\bigl(1 + \rho_u + \textstyle\sum_i \rho_i\bigr) \\
             &= H + \tfrac{D}{4}T_w + \textstyle\sum_i \tfrac{K_i}{12T_i}T_w^2 .
\end{aligned}}
\label{eq:hhat_expansion}
\end{equation}
\rev{Each correction to $H$ is $H$ times the energy fraction delivered by that response on the initial trajectory. Equations~\eqref{eq:ramp}--\eqref{eq:hhat_expansion} and the closed form in~\eqref{eq:hhat_general} require linear, unsaturated delivery and windows before the nadir that satisfy the alignment condition stated after~\eqref{eq:hhat_def}. Equation~\eqref{eq:hhat_expansion} rests in addition on the initial trajectory and on the first-order form of~\eqref{eq:hhat_identity}.}

\rev{For use in planning, the inputs are those a system operator holds for its own system---inertia constants and ratings of the units in service, governor droops and time constants, converter gains and delays, or equivalents where a control chain is reduced to a lag and the output, for a contingency $\Delta P$ and the window $T_w$ of the measuring device, is the inertia the device will report and the RoCoF it will show, $\Delta P/2\hat{H}(T_w)$ in per-unit frequency per second. Section~\ref{sec:results_ieee9} compares the response-model predictions, computed from the parameter set of Table~\ref{tab:model_inputs} by the evaluation methods specified there, with the measured windowed estimates, and, where the response powers are exported, closes the energy identity \eqref{eq:hhat_identity} on them; the two describe the same events.}

\section{Results}
\label{sec:results}

\rev{The IEEE 9-bus results compare the predictions of Section~\ref{sec:framework} with RMS measurements. The reduced Greek system then shows the same window dependence on a larger network and across levels of converter penetration.}

\subsection{IEEE 9-Bus System Results}
\label{sec:results_ieee9}

\subsubsection{Test System and Methods}

\rev{The IEEE 9-bus benchmark was simulated in RMS mode in three configurations. The baseline event is a step increase of Load A at bus 5 (+62.5\,MW); the sensitivity study of Section~\ref{sec:sensitivity} varies its size. All configurations carry the same governor set.}
\begin{enumerate}
\item \textbf{Base Case (Scenario A):} the original three synchronous generators (Table~\ref{tab:sg_constants}).
\item \rev{\textbf{GFL Case (Scenario B):} the machine at bus 3 replaced by a 125\,MVA WECC aggregated positive sequence dynamic model dispatched at 85\,MW, with frequency droop ($K_f = 20$ on the converter base) behind a $\pm 50$\,mHz deadband, a 40\,ms frequency-measurement delay and a power-control loop, and a RoCoF-emulation branch sized at $2H_{G3}$ on a 0.1\,s filter.}
\item \rev{\textbf{GFM Case (Scenario C):} the same machine replaced by a 125\,MVA virtual synchronous machine (VSM) dispatched at 85\,MW, implementing \eqref{eq:vsg_control} with $T_a = 2H_v = 10$\,s ($6.25$\,s on $S_B$) and unfiltered damping $D_p = 58$ on the converter base.}
\end{enumerate}
\rev{Tables~\ref{tab:sg_constants} and~\ref{tab:model_inputs} give the machine data and response-model parameters.}

\paragraph{Machine damping and load relief}
\rev{The undelayed aggregate $D$ contains machine damping and, in Scenario C, GFM damping. The constant-impedance loads reduce the effective deficit below the commanded step; the measured first-sample electrical deficit is 55.4\,MW in the SG base case against the commanded 62.5\,MW, and the deficit is referred to that step, so the load relief at the event instant is excluded from the disturbance rather than counted as a response. The further load-side change inside the window, measured on the governed and ungoverned SG records, is below 0.7\% of the disturbance energy at every window, so no separate load-damping term is included in the response model.}

\paragraph{Parameter values and model reduction}
\label{sec:provenance}
\rev{All parameters use $S_B=100$~MVA and $f_0=50$~Hz. One parameter set is used across windows for each configuration. It combines the declared settings with the governor equivalent estimated on the SG base record and the GFL power-loop equivalent $T_f$ assessed on the GFL records; the comparisons are retrospective. The adopted effective deficits are $55.9/\allowbreak 53.2/\allowbreak 55.8$\,MW for Scenarios A/B/C; each event-size run uses its own measured deficit.}

\paragraph{Evaluation of the model for the studied cases}
\label{sec:evaluation}
\rev{Measured estimates are obtained from the RMS centre-of-inertia (COI) frequency using \eqref{eq:rocof_measured} and~\eqref{eq:hhat_def}. Predicted estimates use the same estimator on the trajectory of \eqref{eq:model_td}, with the parameters of Table~\ref{tab:model_inputs} and the same event deficit. Every configuration is evaluated with the closed form~\eqref{eq:hhat_general}. In the GFL cases the 40\,ms measurement delay of the simulated controller (100\,ms in the delay variant) enters as the first-order lag $\theta$ of~\eqref{eq:model_td}, and its $\pm 50$\,mHz droop deadband is not represented. To size the two simplifications, \eqref{eq:model_td} was also stepped forward in time, first with the exact 40\,ms (100\,ms) dead time in place of the lag, which changes the predictions by at most 0.14\,s (0.40\,s in the delay variant at 300\,ms), and then with the deadband ($\varepsilon = 0.001$~pu) applied to the measured drop: the measured drop crosses the deadband 135\,ms after the event (182\,ms in the delay variant), so the droop, which in the linear model starts delivering at once, is held back until then; this lowers the 500\,ms predictions of the four GFL runs by 0.7--3.0\,s, most at the highest droop gain, and the 100\,ms predictions by at most 0.11\,s. With the deadband retained the model lies within 2.4\% of every GFL measurement, so the deadband accounts for 75--83\% of the 500\,ms excess and the remainder, 0.6--1.5\%, is unattributed model error. Table~\ref{tab:closedform_bracket} compares the closed-form predictions with the RMS estimates.}

\begin{table}[htbp]
\centering
\footnotesize
\color{blue}
\caption{Synchronous machine data and kinetic energies (declared).}
\label{tab:sg_constants}
\begin{tabular}{lcccccc}
\hline
 & \shortstack{$S_{gn}$\\(MVA)} & $\cos\varphi$ & \shortstack{$P_{gn}$\\(MW)} & \shortstack{$H$\\(s)} & \shortstack{$E$\\(MW\,s)} & \shortstack{$P_0$\\(MW)} \\
\hline
G1 & 512 & 0.90 & 460.8  & 2.6312 & 1347.2 & 71.65 \\
G2 & 270 & 0.85 & 229.5  & 4.1296 & 1115.0 & 163.0 \\
G3 & 125 & 0.85 & 106.25 & 4.768  & 596.0  & 85.0  \\
\hline
\end{tabular}
\vspace{2pt}

\raggedright\footnotesize $H$ on the machine rating; $E = H S_{gn}$. System $H=\sum_i E_i/S_B = 30.58$~s on $S_B=100$~MVA for three machines; $24.62$~s for G1--G2. Loads A/B/C $=125/\allowbreak 90/\allowbreak 100$~MW, constant impedance.
\end{table}

\begin{table*}[htbp]
\centering
\footnotesize
\color{blue}
\caption{Response-model inputs for the three base cases, on $S_B=100$~MVA and $f_0=50$~Hz (Section~\ref{sec:provenance}).}
\label{tab:model_inputs}
\begin{tabular}{l c c c}
\hline
\textbf{Parameter} & \textbf{A (SG)} & \textbf{B (GFL)} & \textbf{C (GFM)} \\
\hline
$H$ (s) & 30.58 & 24.62 & 30.87 \\
Machine damping (pu/pu) & 15.93 & 13.81 & 13.81 \\
GFM damping $D_p$ (pu/pu) & --- & --- & 72.5 \\
Undelayed aggregate $D$ (pu/pu) & 15.93 & 13.81 & 86.31 \\
Governor $K_g$ (pu/pu) & 622.6 & 596.1 & 596.1 \\
Equivalent governor branches & \multicolumn{3}{c}{$0.256K_g$ at $T_g{=}0.402$~s; $0.744K_g$ at $T_R{=}8.58$~s} \\
GFL droop $K_f$ (pu/pu) & --- & 25 & --- \\
GFL measurement delay $\theta$ (as first-order lag), lag $T_f$ (s) & --- & 0.040, 0.14 & --- \\
GFL RoCoF-emulation gain $2H_{GFL}$ (pu\,s), filter $T_r$ (s) & --- & 11.92, 0.1 & --- \\
\hline
\end{tabular}
\vspace{2pt}

\end{table*}

\paragraph{Measurement chain}
\rev{The reported measurements use the rotor-speed COI frequency: an unfiltered single-interval slope and the 100, 300 and 500\,ms trailing averages of~\eqref{eq:rocof_measured}. Each estimate uses the corresponding run's effective deficit.}

\subsubsection{Base Case Response (All Synchronous Generators)}

\begin{figure}[!tbp]
    \centering
    \includegraphics[width=\linewidth]{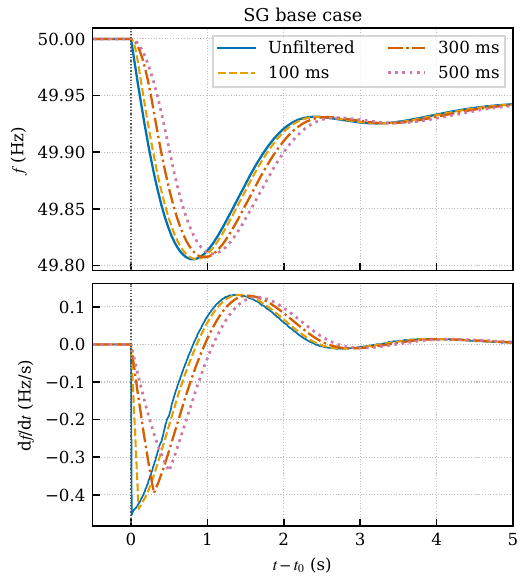}
    \caption{Frequency and RoCoF perception for SG-only case under different measurement windows. \rev{Frequency is the centre of inertia computed from rotor speeds; time is counted from the event instant $t_0$.}}
    \label{fig:freq-sim}
\end{figure}

\rev{For the SG base case, the closed form predicts $31.35/\allowbreak 35.00/\allowbreak 41.80$\,s at 100/300/500\,ms, compared with $31.96/\allowbreak 35.31/\allowbreak 41.47$\,s from the RMS record (Table~\ref{tab:closedform_bracket}). The maximum deviation is 1.9\%. The unfiltered estimate is $30.80$\,s against the reference $30.58$\,s, consistent with~\eqref{eq:rocof0}. Figure~\ref{fig:freq-sim} shows the corresponding frequency and RoCoF traces.}

\rev{The comparison uses COI frequency throughout. Applied to the electrical frequency of single buses of the same record, the estimator spreads about $\pm 7\%$ around the bus mean at 100\,ms, $\pm 1.6\%$ at 300\,ms and $\pm 4\%$ at 500\,ms, and gives no usable unfiltered value because the phase transient at the event dominates the single-interval slope; under the frequency-divider approximation a bus frequency is a network-dependent weighted combination of the machine frequencies \cite{Milano2017Divider}, so the COI chain is used to remove the dependence on the measurement location, while the dependence on the window remains.}

\subsubsection{Grid-Following Inverter Response}

\begin{figure}[!tbp]
\centering
\includegraphics[width=\linewidth]{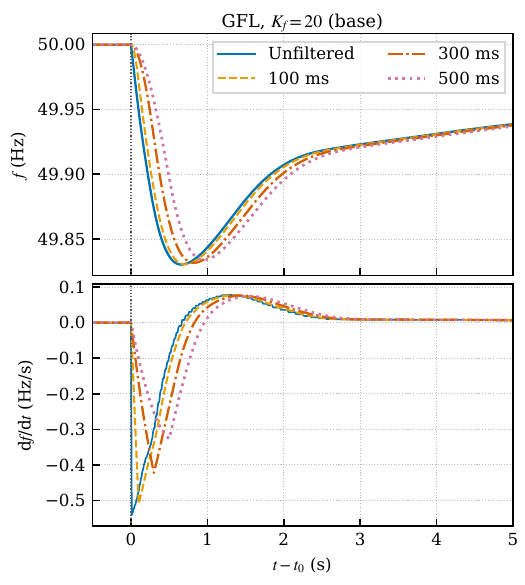}
\caption{Frequency and RoCoF perception for GFL inverter case under different measurement windows. \rev{Frequency is the centre of inertia computed from rotor speeds; time is counted from the event instant $t_0$.}}
\label{fig:gfl_inertia}
\end{figure}

\rev{For the GFL base case, the closed form gives $25.59/\allowbreak 31.60/\allowbreak 42.54$\,s, compared with $26.11/\allowbreak 31.53/\allowbreak 40.78$\,s from the RMS record (Table~\ref{tab:closedform_bracket}, Fig.~\ref{fig:gfl_inertia}); the 500\,ms prediction lies 4.3\% above the measurement, largely because the model carries no droop deadband (Section~\ref{sec:evaluation}). The unfiltered estimate is $24.62$\,s, equal to the synchronous reference inertia, as~\eqref{eq:rocof0} predicts for delayed emulation. At 500\,ms, the measured estimate reaches $1.66$ times this reference; the converter support later withdraws when frequency returns inside its deadband.}

\subsubsection{Grid-Forming Inverter Response}

\begin{figure}[!tbp]
\centering
\includegraphics[width=\linewidth]{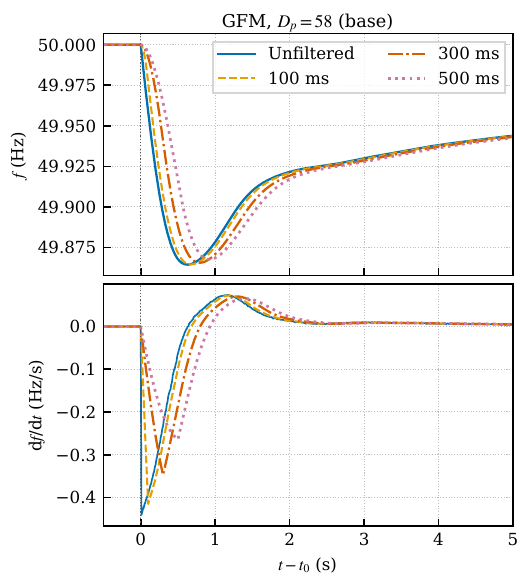}
\caption{Frequency and RoCoF perception for GFM inverter case under different measurement windows. \rev{Frequency is the centre of inertia computed from rotor speeds; time is counted from the event instant $t_0$.}}
\label{fig:gfm_inertia}
\end{figure}

\rev{For the GFM base case, the closed form gives $33.45/\allowbreak 41.37/\allowbreak 53.96$\,s, compared with $33.59/\allowbreak 40.13/\allowbreak 52.76$\,s from the RMS record (Table~\ref{tab:closedform_bracket}, Fig.~\ref{fig:gfm_inertia}), within 3.1\%. The unfiltered estimate is $31.64$\,s against the reference $30.87$\,s, which includes the programmed virtual inertia. The measured 500\,ms estimate is $1.71$ times the reference, consistent with the governor and damping response delivered inside the window.}

\subsubsection{\rev{Comparison Across Configurations}}

\begin{table}[htbp]
\centering
\footnotesize
\color{blue}
\caption{Windowed inertia estimate $\hat{H}(T_w)$ against the reference inertia $H$ for the three IEEE 9-bus base cases (s).}
\label{tab:inertia_9bus}
\begin{tabular}{lccccc}
\hline
\textbf{Case} & $H$ (ref.) & Unfilt. & 100\,ms & 300\,ms & 500\,ms \\
\hline
SG (A)  & 30.58 & 30.80 & 31.96 & 35.31 & 41.47 \\
GFM (C) & 30.87 & 31.64 & 33.59 & 40.13 & 52.76 \\
GFL (B) & 24.62 & 24.62 & 26.11 & 31.53 & 40.78 \\
\hline
\end{tabular}
\end{table}

\rev{Table~\ref{tab:inertia_9bus} shows that the unfiltered estimates agree with the reference inertias within 2.5\%. At 500\,ms, the GFL and SG estimates differ by only 1.7\%, despite the GFL system having 20\% less synchronous inertia.}

\subsubsection{\rev{Agreement with the Response Model}}
\label{sec:validation}

\rev{Table~\ref{tab:closedform_bracket} places the predictions and RMS measurements side by side. Deviations are within 1.9\% for the SG base and ungoverned cases, 3.1\% for the two GFM cases and 2.4\% for the four GFL cases at 100 and 300\,ms; at 500\,ms the linear model lies 2.3--8.7\% above the GFL measurements, increasingly with the droop gain, most of which is the effect of the deadband it does not carry (Section~\ref{sec:evaluation}). The reduced-governor-gain case deviates by 5.5\%. The event-size comparisons are reported separately in Table~\ref{tab:sensitivity_event_size}. The governor equivalent was fitted on the SG base record, so that case and the ungoverned control are within-record checks; the other nine runs use the same equivalent transferred.}

\rev{The energy identity \eqref{eq:hhat_identity} is closed separately on the measured response powers. In the SG case, where turbine power, machine damping and load relief are all exported, the inertia reconstructed from the measured shares is within $0.7\%$ of the reference; in the GFM cases, with $H$ taken as the machine value (the exported converter power contains its inertial part) and the estimated governor equivalent standing in for the unexported turbine power, within $1.2\%$; in the GFL case a residual of $2\%$ of the disturbance energy remains, not attributed because the load relief was not exported for that run. The converter shares in these closures are measured from the converter's own power, so they are not mixed with the load response.}

\paragraph{Model-based correction as a consistency check}
\label{sec:mitigation}
\rev{The identity also provides the correction $H = \hat{H}(T_w)\,(1-\rho(T_w))$. With $\rho$ from the response model this is a consistency check between measured and modelled estimates, not an independent identification of $H$; it agrees with $H$ within 3.0\% for the SG and GFM base cases at every window, including the largest observed bias ($1.71\times$, GFM at 500\,ms), within 4.1\% for the GFL base case (at 500\,ms, where the model carries no deadband) and within 5.8\% for the reduced-droop run. For the studied models and a 5\% bias criterion, uncorrected estimation is confined to $T_w \lesssim 150$\,ms in the studied SG case and $T_w \lesssim 60$\,ms with the studied GFM damping; beyond that either the window is reported with the estimate or the correction is applied.}

\begin{table*}[t]
\centering
\small
\color{blue}
\renewcommand{\arraystretch}{1.15}
\setlength{\tabcolsep}{10pt}
\caption{Predicted and RMS-measured windowed inertia estimates (s). Predictions from the closed form~\eqref{eq:hhat_general} with the inputs of Table~\ref{tab:model_inputs}; the GFL predictions carry no droop deadband (Section~\ref{sec:evaluation}).}
\label{tab:closedform_bracket}
\begin{tabular}{lrrrrrr}
\hline
 & \multicolumn{2}{c}{100\,ms} & \multicolumn{2}{c}{300\,ms} & \multicolumn{2}{c}{500\,ms} \\
\cline{2-7}
\textbf{Case} & \textbf{Predicted} & \textbf{RMS} & \textbf{Predicted} & \textbf{RMS} & \textbf{Predicted} & \textbf{RMS} \\
\hline
SG ungoverned & 31.39 & 31.83 & 33.03 & 33.39 & 34.74 & 35.03 \\
SG (A) & 31.35 & 31.96 & 35.00 & 35.31 & 41.80 & 41.47 \\
SG (A), $K_g{=}159$ & 31.07 & 31.89 & 32.57 & 33.92 & 34.69 & 36.71 \\
GFL (B), $K_f{=}20$ & 25.59 & 26.11 & 31.60 & 31.53 & 42.54 & 40.78 \\
GFL (B), $K_f{=}10$ & 25.56 & 26.11 & 31.18 & 31.44 & 41.10 & 40.18 \\
GFL (B), $K_f{=}40$ & 25.64 & 26.11 & 32.46 & 31.70 & 45.61 & 41.96 \\
GFL (B), $T_d{=}100$\,ms & 25.45 & 25.82 & 30.72 & 30.98 & 41.33 & 39.62 \\
GFM (C), $D_p{=}58$ & 33.45 & 33.59 & 41.37 & 40.13 & 53.96 & 52.76 \\
GFM (C), $D_p{=}29$ & 32.50 & 33.51 & 38.08 & 38.47 & 47.37 & 47.09 \\
\hline
\end{tabular}
\end{table*}

\subsubsection{\rev{Sensitivity to Event Size and Control Parameters}}
\label{sec:sensitivity}

\begin{figure}[!tbp]
\centering
\includegraphics[width=\linewidth]{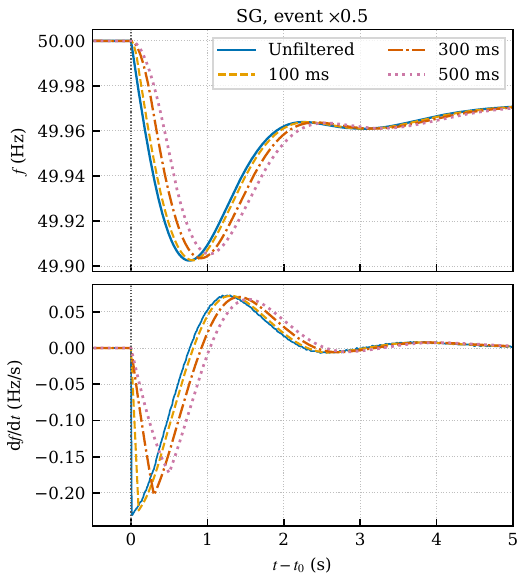}
\caption{\rev{Frequency and RoCoF perception, SG base case with event $\times 0.5$ (+31.25\,MW commanded), under different measurement windows. Frequency is the centre of inertia computed from rotor speeds; time is counted from the event instant $t_0$.}}
\label{fig:s5a}
\end{figure}

\begin{figure}[!tbp]
\centering
\includegraphics[width=\linewidth]{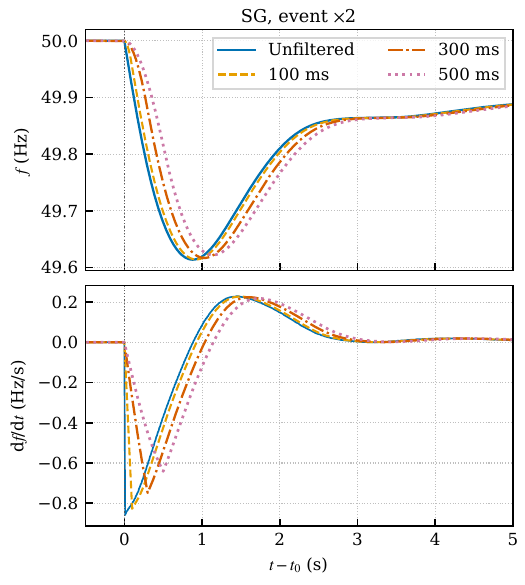}
\caption{\rev{Frequency and RoCoF perception, SG base case with event $\times 2$ (+125\,MW commanded), under different measurement windows.}}
\label{fig:s5b}
\end{figure}

\rev{\emph{Event size.} Halving and doubling the SG load step changes the measured effective deficit from 28.2 to 105.0\,MW (Table~\ref{tab:sensitivity_event_size}, Figs.~\ref{fig:s5a}--\ref{fig:s5b}). The closed-form estimate is independent of disturbance size for this linear model, and all measured estimates remain within 2.3\% of its prediction.}

\rev{\emph{Governor gain.} At 500\,ms, the predictions for $K_g=0/\allowbreak 159/\allowbreak 623$ are $34.74/\allowbreak 34.69/\allowbreak 41.80$\,s, compared with $35.03/\allowbreak 36.71/\allowbreak 41.47$\,s measured. The reduced-gain case deviates by up to 5.5\%; the shortfall is associated with the transferred governor equivalent, whose 500\,ms response share is 0.058 against 0.101 measured. No parameter was re-estimated for this case.}

\begin{figure}[!tbp]
\centering
\includegraphics[width=\linewidth]{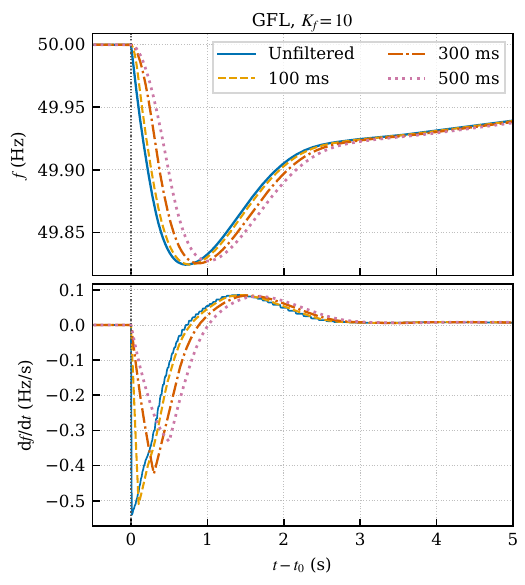}
\caption{\rev{Frequency and RoCoF perception, GFL case with $K_f=10$, under different measurement windows.}}
\label{fig:gfl_kf10}
\end{figure}

\begin{figure}[!tbp]
\centering
\includegraphics[width=\linewidth]{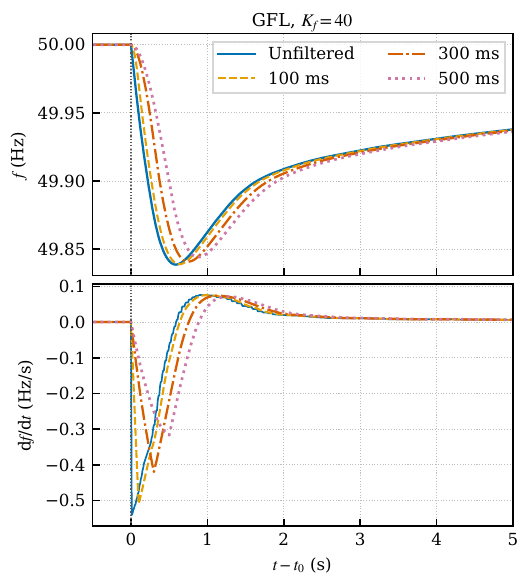}
\caption{\rev{Frequency and RoCoF perception, GFL case with $K_f=40$, under different measurement windows.}}
\label{fig:gfl_kf40}
\end{figure}

\begin{figure}[!tbp]
\centering
\includegraphics[width=\linewidth]{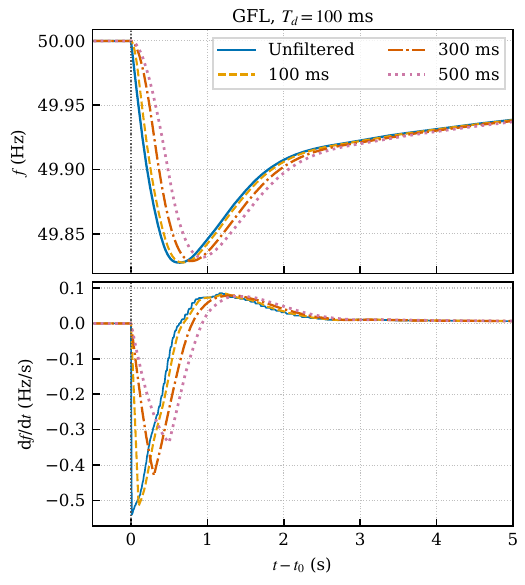}
\caption{\rev{Frequency and RoCoF perception, GFL case with measurement delay $T_d=100$\,ms ($K_f=20$), under different measurement windows.}}
\label{fig:gfl_td100}
\end{figure}

\rev{\emph{GFL controls.} The model follows the direction of every GFL variation (Table~\ref{tab:closedform_bracket}, Figs.~\ref{fig:gfl_kf10}--\ref{fig:gfl_td100}). Increasing $K_f$ from 10 to 40 raises the measured 500\,ms estimate from 40.18 to 41.96\,s and the predicted one from 41.10 to 45.61\,s: the linear model over-states the sensitivity to the droop gain because, without the deadband, the droop begins to deliver at the event instead of 135\,ms after it, and the excess grows with $K_f$; with the deadband retained the three predictions are 40.41, 41.11 and 42.57\,s (Section~\ref{sec:evaluation}). With a 100\,ms delay the 100\,ms prediction falls from 25.59 to 25.45\,s as the measurement falls from 26.11 to 25.82\,s, and the 500\,ms prediction is 41.33\,s against 39.62\,s measured.}

\begin{figure}[!tbp]
\centering
\includegraphics[width=\linewidth]{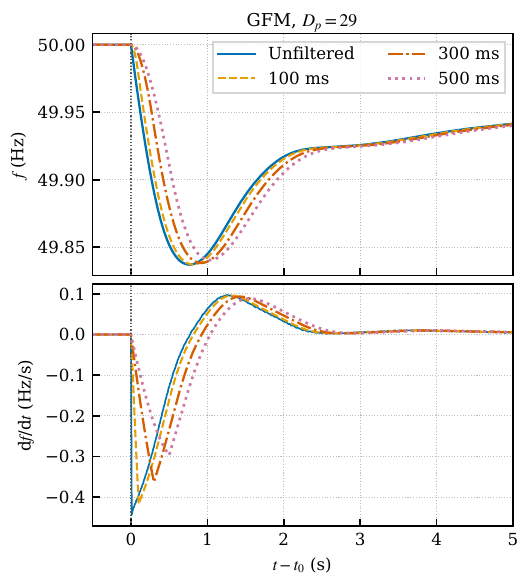}
\caption{\rev{Frequency and RoCoF perception, GFM case with $D_p=29$ ($T_a=10$\,s), under different measurement windows.}}
\label{fig:gfm_dp29}
\end{figure}

\rev{\emph{GFM damping.} Halving $D_p$ from 58 to 29 reduces the predicted 500\,ms estimate from 53.96 to 47.37\,s, compared with 52.76 to 47.09\,s measured (Fig.~\ref{fig:gfm_dp29}). Both damping cases agree with the model within 3.1\% across the windows; their unfiltered estimates change by less than 0.5\%.}

\begin{table*}[htbp]
\centering
\footnotesize
\color{blue}
\caption{Event-size sensitivity of the windowed inertia estimate, SG base case, three disturbance magnitudes, with the model value.}
\label{tab:sensitivity_event_size}
\begin{tabular}{lcccccc}
\hline
 & & \multicolumn{4}{c}{$\hat{H}(T_w)$ (s)} & \\
\textbf{Event} & $\Delta P_{\mathrm{eff}}$ (MW) & \textbf{Unfilt.} & \textbf{100\,ms} & \textbf{300\,ms} & \textbf{500\,ms} & Nadir (mHz) \\
\hline
$\times 0.5$ (+31.25\,MW) & 28.2  & 30.58 & 31.44 & 34.70 & 40.86 & $-97.6$ \\
$\times 1$ (+62.5\,MW)    & 55.9  & 30.80 & 31.96 & 35.31 & 41.47 & $-194.6$ \\
$\times 2$ (+125\,MW)     & 105.0 & 30.58 & 31.73 & 35.08 & 40.94 & $-386.3$ \\
Model, any size          & ---   & 30.63 & 31.35 & 35.00 & 41.80 & --- \\
\hline
\end{tabular}
\vspace{2pt}

\raggedright\footnotesize Commanded step in parentheses; $\Delta P_{\mathrm{eff}}$ is measured per run. The model row is the closed form of \eqref{eq:model_td} with the Scenario-A constants of Table~\ref{tab:model_inputs}, its unfiltered entry evaluated with the same 10\,ms single interval as the measured column (the $T_w \to 0$ limit is $H = 30.58$\,s); the linear model's estimate does not depend on the event size, and the measured values lie within 2.3\% of it.
\end{table*}

\subsubsection{\rev{Reference PMU Filter Results}}
\label{sec:pmu_reference}

\rev{Table~\ref{tab:pmu_reference} compares estimates from the IEC/IEEE 60255-118-1:2018 Annex~D reference filters \cite{iec60255_2018} with the moving-average estimates for the same COI records. The P-class triangular and M-class sinc$\times$Hamming filters use the standard's phase-difference calculations and reporting rates; each range covers all reporting phases.}

\rev{The P-class and 10\,fps M-class estimates match moving-average estimates at 37--82\,ms and 54--244\,ms, respectively. At the phases marked in the table, the 50 and 25\,fps M-class filters produce estimates below the unfiltered value because their signed kernels overshoot the RoCoF peak; no positive moving-average window reproduces those values. The comparison therefore matches reported inertia values where possible, without implying equivalence between the filters.}

\begin{table*}[htbp]
\centering
\footnotesize
\color{blue}
\caption{Inertia estimates obtained with the selected IEC/IEEE 60255-118-1 reference filters applied to the COI records: range over all reporting phases (s), and the moving-average windows giving the same estimate where one exists.}
\label{tab:pmu_reference}
\begin{tabular}{lccccc}
\hline
 & & \multicolumn{3}{c}{\textbf{Estimate range over reporting phases (s)}} & \\
\textbf{Reference model} & ROCOF resp.\ limit & \textbf{SG} & \textbf{GFL} & \textbf{GFM} & equiv.\ $T_w$ \\
\hline
P class, 50\,fps  & 120\,ms & 31.32--31.77 & 24.99--25.51 & 32.31--32.95 & 37--82\,ms, all phases \\
M class, 50\,fps  & 280\,ms & 29.83--30.73 & 23.89--24.75 & 30.88--32.01 & none at 16/16, 14/16, 12/16 phases$^{\ast}$ \\
M class, 25\,fps  & 560\,ms & 29.80--31.24 & 24.22--25.68 & 31.30--33.07 & none at 25/32, 16/32, 13/32 phases$^{\ast}$ \\
M class, 10\,fps  & 1.4\,s  & 31.49--33.86 & 27.39--29.87 & 34.36--37.56 & 54--244\,ms, all phases \\
\hline
Reference inertia $H$ & --- & 30.58 & 24.62 & 30.87 & --- \\
Unfiltered estimate (measured) & --- & 30.80 & 24.62 & 31.64 & $T_w\!\to\!0$ \\
\hline
\end{tabular}
\vspace{2pt}

\raggedright\footnotesize $^{\ast}$At the phases counted (SG, GFL, GFM) the estimate lies below the same record's unfiltered estimate and no positive moving-average window gives it; at the other phases a window of 10--80\,ms does. Response-time limits from Table~9 of the standard.
\end{table*}

\subsection{Reduced Greek System Results}
\label{sec:results_greek}

\rev{The reduced Greek transmission system provides a larger-network application of the windowed estimator across three levels of converter penetration.}

\subsubsection{Greek System Model Description}

The model contains 218 equivalent buses, representing the 400\,kV backbone and key 150\,kV substations, and 61 equivalent synchronous generators. It includes thermal, hydro, wind and solar generation, with neighbouring systems represented by equivalent AC ties. In the baseline, 45 machines belong to the connected system.

\subsubsection{Scenario Definition}

Five dispatch scenarios were defined: G1 is the baseline without SG replacement; G2 and G3 replace synchronous generation with IBRs supplying 10\% and 25\% of total generation, respectively. In subcases ``a'', 10\% of IBR capacity operates in GFM mode and 90\% in GFL mode; in subcases ``b'', the shares are 25\% and 75\%. The results below cover G1, G2b and G3b.

\subsubsection{\rev{Windowed Estimates on the Greek System}}

\rev{The COI frequency, windowed RoCoF and inertia estimate were calculated from the exported machine records using the procedure of Section~\ref{sec:results_ieee9} on a 10\,ms grid. At $t=200$\,s, the largest generating equivalent and the small equivalents connected to its switched node are disconnected. The remaining $39/\allowbreak 34/\allowbreak 26$ machines define the reference inertia in G1/G2b/G3b, $69.2/\allowbreak 62.3/\allowbreak 45.5$\,GW\,s; the effective deficit, measured from the first post-event electrical-power sample, is $2229/\allowbreak 2424/\allowbreak 2420$\,MW. Every machine entering the COI is included, including the REI and interconnection equivalents with inertia constants of 4 and 7\,s, respectively.}

\rev{Only measured quantities are reported for this system: the response-model prediction is not evaluated, because the load and network response of this model is outside the reduced model of Section~\ref{sec:framework}, and the converter powers were not exported, so the response energy is not attributed to its sources.}

\begin{figure}[!tbp]
\centering
\includegraphics[width=\linewidth]{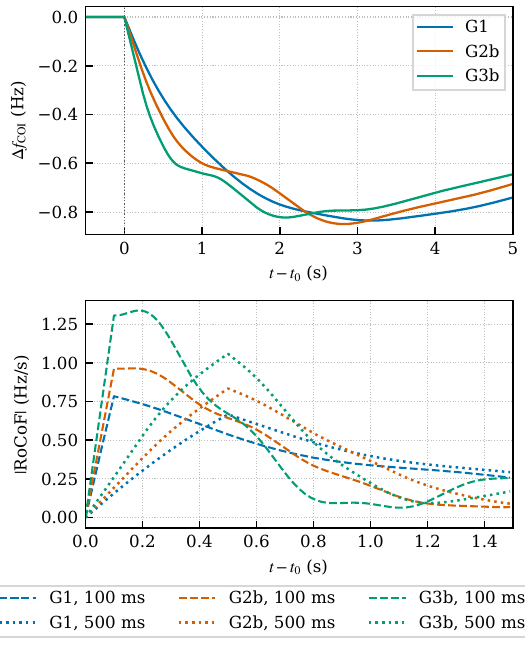}
\caption{\rev{Reduced Greek system, loss of the largest generating equivalent at $t_0$: deviation of the centre-of-inertia frequency of the connected machines (top) and magnitude of the windowed RoCoF \eqref{eq:rocof_measured} for the 100 and 500\,ms windows (bottom), scenarios G1, G2b and G3b.}}
\label{fig:greek_coi}
\end{figure}

\rev{Fig.~\ref{fig:greek_coi} shows the centre-of-inertia frequency and the windowed RoCoF of the three scenarios. The unfiltered RoCoF is $0.80/\allowbreak 0.97/\allowbreak 1.35$\,Hz/s in G1/G2b/G3b; the 100, 300 and 500\,ms windows lower it to $0.79/\allowbreak 0.73/\allowbreak 0.67$\,Hz/s in G1, $0.97/\allowbreak 0.93/\allowbreak 0.83$\,Hz/s in G2b and $1.34/\allowbreak 1.26/\allowbreak 1.06$\,Hz/s in G3b. The estimates \eqref{eq:hhat_def} that follow from these values are $69.3/\allowbreak 62.4/\allowbreak 44.7$\,GW\,s unfiltered, within 0.3\% of the connected-machine inertia for G1 and G2b and within 1.7\% for G3b, and $71.0/\allowbreak 76.3/\allowbreak 83.6$\,GW\,s (G1), $62.8/\allowbreak 65.0/\allowbreak 72.6$\,GW\,s (G2b) and $45.2/\allowbreak 47.9/\allowbreak 57.1$\,GW\,s (G3b) at 100/300/500\,ms. For three of these values (G2b at 100\,ms, G3b unfiltered and at 100\,ms) the maximising window of \eqref{eq:hhat_def} ends $0.15$--$0.19$\,s after the event; the event-aligned values are $63.0$, $45.6$ and $46.3$\,GW\,s. At 500\,ms the estimates thus exceed the inertias by 17--26\%; the identity \eqref{eq:hhat_identity} relates this excess to the aggregate response delivered inside the window, without identifying its sources. Read across the scenarios, the window changes what the measurement says about the system: the 500\,ms window lowers the measured RoCoF below its unfiltered value by 17\% in G1 (0.80 to 0.67\,Hz/s), 14\% in G2b (0.97 to 0.83\,Hz/s) and 22\% in G3b (1.35 to 1.06\,Hz/s), and in inertia terms G2b, with 10\% less synchronous inertia than G1, reports $72.6$\,GW\,s at 500\,ms, more than G1's actual $69.2$\,GW\,s, while G3b, with 34\% less, reports $57.1$\,GW\,s, 83\% of G1's inertia. The unfiltered estimates show the actual proportions ($62.4$ and $44.7$\,GW\,s against $69.3$); the 500\,ms values do not. At 500\,ms, all three scenarios therefore report estimates above their connected-machine inertia, and the response delivered inside the window stands in, in the reported value, for part of the synchronous inertia removed, while substantial differences between scenarios remain. The earlier version of this section reported the peak RoCoF of the electrical frequency at a single bus; those unfiltered peaks were 2.3--2.7 times the centre-of-inertia values found here, as the location dependence measured on the 9-bus system leads one to expect, and that table is not retained.}

\section{Discussion}

\subsection{Impact of Measurement Window on Perceived Response}

\rev{Two findings of Section~\ref{sec:results} carry the discussion. First, the windowed estimate grows with the window for every interface technology, by comparable factors at 500\,ms, and in doing so blurs the differences between systems: the GFL scenario, with 20\% less synchronous inertia than the SG base case, reports an estimate only 1.7\% lower, while the GFM scenario reads the highest value although its reference inertia is close to the SG case's. Second, the unfiltered estimate returned the reference inertia within 2.5\% in all eleven reported 9-bus runs, across three governor configurations, three event sizes and three interface technologies, and within 2\% in the three Greek-system scenarios; because the disturbance estimate uses the machine inertias in most runs this is a consistency check of the accounting rather than an independent measurement of $H$, but within that scope it is the reference against which windowed estimates are read and, for the 9-bus cases, interpreted through the response shares. The simulations are RMS with the full controller models of the units; electromagnetic transients, control-bandwidth limits and measurement noise are outside the study, and no emergency scheme acts in the studied windows: the 9-bus nadirs lie between $-98$ and $-386$\,mHz (Table~\ref{tab:sensitivity_event_size}), above under-frequency load-shedding thresholds, and the estimates are read within 500\,ms of the event.}

\subsection{Implications for System Planning and Operation}

The measurement-induced perception of inertia has significant implications for power system planning and operation, consistent with emerging international experience \cite{AEMO_InertiaRequirements_2024,ENTSOE_GFM_2024}:

\textbf{Planning and Markets:} The mix of synchronous and virtual inertia a plan requires, and the services a market can distinguish, depend on the window behind the measurement: responses that look identical at 500\,ms have different dynamics, and a study or scheme setting calibrated to a windowed value is calibrated to a quantity that changes with the window. Performance-based, technology-neutral requirements have been proposed for this reason \cite{Tayyebi2020}; the paragraph below states what the present results add to them. \rev{As an illustration of the calculation, take the Scenario-A parameter set of Table~\ref{tab:model_inputs} as a specified planning condition: for $T_w = 300$\,ms the closed form \eqref{eq:hhat_general} gives $\hat{H} = 35.00$\,s, and for a specified deficit of $55.9$\,MW on the 100\,MVA base the windowed RoCoF a 300\,ms device would show is $f_0 \Delta P/2\hat{H} = 50 \times 0.559/(2 \times 35.00) = 0.40$\,Hz/s, against $0.46$\,Hz/s unfiltered. The example reuses a case that was assessed retrospectively in Section~\ref{sec:results_ieee9}; it illustrates the calculation, not an additional validation.}

\textbf{Inertia Procurement:} \rev{A windowed measurement serves two different purposes, and the framework separates them. When the purpose is to recover the inertia $H$---a requirement stated as a floor on $H$ and checked from measured estimates---the response delivered inside the window must be accounted for: uncorrected, the estimate exceeds $H$ by the factor $1/(1-\sum\rho)$ of \eqref{eq:hhat_identity}, measured here as 1.36--1.71 at 500\,ms, and a requirement checked against such estimates would be met with less inertia than it intends. When the purpose is performance against a windowed RoCoF criterion---a minimum inertia level derived from a RoCoF limit evaluated over a 500\,ms interval (Section~\ref{sec:freq-meas-standards})---the quantity constrained is $\hat{H}(T_w)$ at that interval, not $H$ (Eq.~\eqref{eq:hhat_def}), and every response delivered inside the interval contributes to meeting it, so physical inertia is not the only means of compliance. Within the short-window regime of the planning form \eqref{eq:hhat_expansion}, the increments $D\,T_w/4$ and $K_i T_w^2/12T_i$ are the model's sensitivities of the criterion to an undelayed and a first-order-lag response, and can be used to compare how candidate responses affect the selected metric before any of them is dispatched; they are not universal exchange rates, and determining procurement quantities or a least-cost portfolio additionally requires an optimisation with costs, availability, headroom and the applicable frequency and network constraints, which this study does not formulate. In either use, the requirement should name the window and the classes of response it counts. Two further limits apply. A response that withdraws after the window---the GFL droop of Section~\ref{sec:results_ieee9} returns to its setpoint once the frequency re-enters its deadband---has contributed to the windowed value without remaining available, so a requirement that counts it must also state the holding time it expects. And the framework describes the centre-of-inertia signal: a requirement assessed from a single bus adds the local effects noted in Section~\ref{sec:results_ieee9}, which the model sensitivities do not cover.}

\textbf{Protection Coordination:} RoCoF relay settings act on a filtered signal, and the window of that filter is part of the setting; the SA 2016 and GB 2019 events, in which RoCoF-based protection of embedded generation shaped the outcome (about 500\,MW disconnecting in the latter), illustrate the stake \cite{SA_Event_2016,UK_Event_2019}. The present study did not evaluate protection schemes.

\subsection{Grid Code Evolution}

Future grid codes should incorporate explicit provisions to ensure the consistent and robust assessment of inertia and fast frequency response from inverter-based resources. In particular, the following aspects should be addressed:
\begin{enumerate}
\item Explicitly define measurement and evaluation methodologies for virtual and synthetic inertia, including the selection of time windows and filtering techniques.
\item Adopt technology-neutral, performance-based requirements that specify measurable frequency-domain responses rather than prescriptive control implementations.
\item Recognise, within clearly defined measurement frameworks, the functional equivalence between fast frequency response (FFR) and inertial response.
\item Establish standardised testing and validation procedures that explicitly account for the impact of measurement window length and signal processing on reported inertia metrics.
\end{enumerate}

Recent regulatory developments, such as IEEE Std~2800--2022~\cite{IEEE_2800_2022} and ENTSO-E's emerging grid-forming requirements~\cite{ENTSOE_GFM_2024}, represent important steps toward a harmonised, performance-based approach for integrating inverter-based resources into low-inertia power systems.

\subsection{Empirical Validation Opportunities}

Future validation should \rev{draw on} emerging GFM pilot projects (National Grid ESO's stability pathfinder, AEMO's system strength services) and grid-forming battery storage deployments. TSOs should establish monitoring programs to capture high-resolution frequency data during events for comparison with simulation predictions. \rev{The framework of Section~\ref{sec:framework} makes such validation concrete: for a specified operating condition, disturbance and measurement window, the response model predicts the windowed inertia estimate from its parameter values, some of which, as in this study, may have to be estimated from records rather than read from settings; a measured frequency record provides the comparison estimate; and attribution through the energy identity \eqref{eq:hhat_identity} additionally requires the delivered response-power records---turbine, converter and load---which staged tests or well-instrumented events with a known deficit can supply.}

\section{Conclusion}

This paper has \rev{demonstrated} that the perception of virtual inertia from IBRs is \rev{shaped} by the measurement methodologies employed by TSOs. The widespread use of moving-average frequency and RoCoF filters, originally designed for synchronous-generator-dominated systems, leads to a misclassification of fast frequency response from IBRs as inertial response.

Key findings from this study include:
\begin{itemize}
\item \rev{GFL fast response is re-classified as inertia by the measurement window: the base-case estimate rises from $24.62$\,s (unfiltered, equal to the physical inertia of the remaining machines) to $40.78$\,s at 500\,ms ($1.66\times$). The converter's implemented RoCoF-emulation branch, delivered behind its measurement chain, is unobservable at the disturbance instant---consistent with the analysis that GFL inertia does not constrain initial RoCoF \cite{Ducoin2023GFL}.}
\item \rev{GFM virtual inertia acts as inertia on the actual frequency: the unfiltered estimate returns the reference inertia that includes the programmed coefficient ($31.64$ versus $30.87$\,s, $+2.5\%$; the two remaining machines alone would give $24.62$\,s). The GFM windowed estimate nevertheless inflates the most ($1.71\times$ at 500\,ms), because the large damping gain acting on the actual frequency and the governor response delivered through the turbine dynamics accumulate inside long windows.}
\item \rev{Long measurement windows blur the differences between systems: at 500\,ms the GFL-based system, with 20\% less synchronous inertia than the SG system, reports an estimate only 1.7\% lower ($40.8$ against $41.5$\,s), while the GFM-based system reads $52.8$\,s; on the reduced Greek system the scenario with 10\% less synchronous inertia reports, at 500\,ms, more than the all-synchronous scenario's actual inertia ($72.6$ against $69.2$\,GW\,s), and the scenario with 34\% less reports 83\% of it.}
\item \rev{The windowed inertia estimate is predictable by one rule---the estimator applied to a modelled trajectory, in closed form for one linear response model: across the eleven reported runs the model, from declared settings, one governor equivalent fitted on the SG base record and one equivalent lag for the GFL power-control chain, reproduces the measured estimates within 3.1\% (5.5\% for the reduced-droop run), except at 500\,ms in the GFL runs, where the linear model lies 2.3--8.7\% above the measurement and the neglected droop deadband accounts for most of the excess, in evaluations that include two within-record consistency checks and nine runs with the parameters transferred; the identity $\hat{H}(T_w)=H/(1-\rho(T_w))$ reconstructs the reference inertia within $0.1$--$1.2\%$ from measured response powers in the SG and GFM 9-bus cases (with the estimated governor response standing in for the unexported turbine power in the GFM cases) while the GFL closure leaves an unattributed residual of 2\% of the disturbance energy; and the model-based correction, a consistency check whose response fractions depend on the reference inertia, agrees with $H$ within 3.0\% at every window for the SG and GFM base cases, 4.1\% for the GFL base case and 5.8\% for the reduced-droop run.}
\end{itemize}

From the measurement's point of view, the operative distinction is not between ``true'' and ``virtual'' inertia but between response delivered inside the window and response delivered outside it; services assessed with inherited measurement chains are valued accordingly, and a consistent definition of frequency support requires the window to be part of the definition. \rev{Two operational statements follow from the studied cases: the window length $T_w$ should be reported alongside every inertia estimate, since the estimate is not defined without it; and, for the models and the 5\% bias criterion used here, estimation should either be confined to short windows---about 150\,ms for the studied synchronous case, about 60\,ms with the studied GFM damping---or be accompanied by the model-based correction of Section~\ref{sec:mitigation}. Whether these thresholds transfer to other systems depends on their response constants, measurement noise and disturbance estimation, which this study does not address.}

\color{black}
\FloatBarrier

\bibliographystyle{IEEEtran}
\bibliography{bibliography}

\end{document}